**Vortex ratchet reversal at fractional matching fields in kagomé-like array with symmetric pinning centers.**

D. Perez de Lara[1,2], A. Alija[3], E. M. Gonzalez[2], M. Velez[3], J. I. Martin[3], J. L. Vicent[1,2]

[1]IMDEA-Nanociencia, Cantoblanco, 28049 Madrid, Spain
[2]Departamento Física de Materiales, Universidad Complutense. 28040 Madrid, Spain
[3]Departamento de Física, Universidad de Oviedo-CINN, 33007 Oviedo, Spain

**Abstract**

Arrays of Ni nanodots embedded in Nb superconducting films have been fabricated by sputtering and electron beam lithography techniques. The arrays are periodic triangular lattices of circular Ni dots arranged in a kagomé-like pattern with broken reflection symmetry. Relevant behaviors are found in the vortex lattice dynamics : i) At values lower than the first integer matching field, several fractional matching fields are present when the vortex lattice moves parallel or perpendicular to the reflection symmetry axis of the array showing a clear anisotropic character in the magnetoresistance curves, ii) injecting an ac current perpendicular to the reflection symmetry axis of the array yields an unidirectional motion of the vortex lattice (ratchet effect) as a result of the interaction between the whole vortex lattice and the asymmetric lattice of dots, iii) increasing the input current amplitudes the ratchet effect changes polarity independently of matching field values. These experimental results can be explained taking into account the vortex lattice density.

1. INTRODUCTION

Superconducting vortex lattice dynamics have been studied for many years from different points of view. One of the main topics has been the behavior of the vortex lattice when it moves on periodic pinning potentials.[1] Superconducting films fabricated on substrates with arrays of periodic non-superconducting nanostructures are the ideal tool to study topics like vortex ratchet.[2] In brief, an ac current applied to the device induces output dc voltage; so that a rectifier effect appears in the device; i.e. vortex lattice driven by alternating forces leads to a net vortex flow. Worth to note that ratchet effect is a very interesting and wide-ranging effect, remarkable in many fields.[3] Two ingredients are needed to achieve ratchet effect, i) the particles, in our case the vortices, must be out of equilibrium, and ii) they have to move on asymmetric potentials. So far we know, different approaches have been used to fabricate asymmetric pinning potentials, for instance asymmetric pinning centers such as triangles,[2] symmetric square pinning centers with different sizes,[4] grading circles,[5] spacing-graded density of pinning centers[6] or arrow-shaped wedges cages.[7]

Other remarkable vortex dynamics effects are related to commensurability between the vortex lattice and the array of periodic pinning potentials. The vortex lattice motion is slowed down and a minimum appears in the magnetoresistance when the density of the pinning centers equals the density of the vortex lattice. The first order or main minimum happens at magnetic field $H_1 = (\Phi_0/S)$, where S is the unit cell area of the pinning array and $\Phi_0 = 20.7$ G $\mu m^2$ is the quantum fluxoid. Upper order matching fields take place at $H_n = n (\Phi_0/S)$, where n >1 is an integer number. Minima in the resistivity can be also observed at fractional matching values $H_f = f (\Phi_0/S)$, being f a non-integer number.[8] The published works have been mainly focused on periodic pinning arrays with triangular, square or rectangular geometry.[1,9] Other regular arrays have been studied too as kagomé [10,11] or honeycomb.[12,13] Even, quasiperiodic arrays generate peculiar matching effects[14] and, recently, the interaction between the vortex lattice and short-range ordered arrays of pinning centers has been explored too.[15]

Different types of vortices could participate in these phenomena: i) vortices placed at the pinning centers, that can be single quantized[16] or multiple quantized[17] vortices, ii) interstitial vortices placed among pinning centers.[18]

In this work, we analyse, vortex lattice matching and ratchet effects, in completely different scenario: symmetric circular dots arranged in a periodic array with broken reflection symmetry, so that the origin of the asymmetric pinning potentials is the geometry of the periodic array itself. Thus, the ratchet effect will be a lattice effect created by the interaction between the whole vortex lattice and the pinning array, rather than by the "single vortex- single pinning center" potential. We will show that, using the same symmetric pinning centers, vortex ratchet effect could be achieved at fractional matching conditions, and, also, the polarity of the vortex ratchet output signal can be tuned by means of the ac input signal strength. This reversal effect is independent of the order of the fractional or integer matching fields. The number of vortices in this system does not play any role in the vortex reversal and the net flow of the vortices is only reversed modifying the strength of the driving forces.

The paper is organized as follows; we begin discussing the asymmetric patterned potential, we continue with the fabrication and magnetotransport measurement techniques. In the next section, we will present the magnetotransport experimental results and their discussion, beginning with matching effects and followed by ratchet experiments. Finally, a summary of the relevant conclusions will close the paper.

## 2. EXPERIMENTAL

A perfect triangular lattice of dots (see sketch in Fig. 1(a)) could be the starting point to fabricate a periodic but asymmetric array. This can be achieved by eliminating dots at selected positions. First of all, the well known kagomé periodic array can be obtained by taking out ¼ of the pinning centers from the triangular lattice (see sketch in Fig. 1(b)). This kagomé lattice can be thought of as a set of triangles pointing up and down, however, it does not show broken reflection symmetry. The next step to generate a periodic asymmetric array can be seen in the sketch of Fig. 1(c) in which 3 out of 9 pinning sites are eliminated. The outcome is a kagomé-like lattice that exhibits a reflection symmetry axis from the base to the tip of the triangles (Y direction in Fig. 1(c)), but not in the perpendicular direction (X direction in Fig. 1(c)). Therefore a periodic pinning potential with broken reflection symmetry can be fabricated using triangular arrays of circular Ni dots as building blocks. The array is composed of two opposite asymmetric regions: an asymmetric triangular-shaped array of 6 dots and an "empty" cage with triangular shape and pointing out in the reversed direction. Both can provide the asymmetric pinning potentials needed for the observation of ratchet effects. Standard electron beam lithography, magnetron sputtering and etching techniques have been used to fabricate the hybrid devices on Si (100) substrates. More experimental details are given in Ref. 19. The required pattern is obtained by electron-beam writer using polymethyl methacrylate resist and lift-off processing. A 100 μm x 100 μm kagomé-like array of Ni dots with a spacing of 400 nm was fabricated on a Si substrate. Ni dot dimensions are 40 nm thickness and 200 nm diameter, i.e. the same as in the triangular array of Ref. 16. Therefore the saturation number for these dots is one vortex per dot.[16] Finally, a 100 nm thick Nb film is deposited by magnetron sputtering on top of the Ni dot array. Optical photolithography and ion etching have been used to define a cross-shaped 40 μm wide bridge to perform magnetotransport measurements. The magnetic field is applied perpendicular to the sample plane. Magnetoresistance is measured with a commercial Helium cryostat. Vortices are driven by dc currents or ac injected currents (frequency = 10 kHz).

## 3. RESULTS AND DISCUSSION

### 3.1 Commensurability in kagomé–like array.

Figure 2 shows the magnetoresistance experimental results in Nb film with embedded kagomé-like array of Ni dots, at T = 7.95 K, when the vortex lattice moves parallel to the reflection axis (Fig. 2(a)) and perpendicular to the reflection axis (Fig. 2(b)). The comparison between both curves shows a clear anisotropy in the transport properties of the kagomé-like array, with much lower values of the low field magnetoresistance for vortex lattice motion along the Y axis of the array. In both plots a clear structure is observed with sharp and shallow minima, as summarized in Table I. The corresponding plots of the magnetic field minima values vs. the order of those minima are shown in the insets of Fig. 2; from the highest negative to the highest positive field values which yield magnetoresistance minima. In spite of the different shapes of the magnetoresistance curves, a similar spacing between minima is obtained in both situations: 16.2 Oe for vortex lattice movement along Y-axis and 16.3 Oe when the lattice moves along X-axis.

Taking into account that the dot periodicity is 400 nm, it is straightforward to calculate the pinning density and to work out the main matching field value; i. e. the magnetic field obtained when the density of the pinning centers equals the density of the vortex lattice. As sketched in Fig. 1 (c), the kagomé-like array is a triangular lattice in which only 6 out of 9 pinning centers remain, so the main matching field ( $H^{n=1}$ ), is:

$$H^{n=1}_{kagomé-like} = \frac{2}{3} H^{n=1}_{triangular} \quad (1)$$

This calculated value is 100 Oe, which is in good agreement with the largest matching field observed. Thus, all the other minima correspond to fractional matching ($H_f$). The positions of the rest of the minima observed in the kagomé-like array can be calculated by eliminating vortices one by one from the 6 vortices which provide the same density that the density of pinning sites. The theoretical positions of these fractional minima are extracted from:

$$H^{f}_{kagomé-like} = f \cdot H^{n=1}_{kagomé-like} = \frac{p}{6} \cdot \frac{2}{3} H^{n=1}_{triangular} = p \cdot \frac{1}{9} H^{n=1}_{triangular} \quad (2)$$

This gives a series of equally spaced minima at $\Delta H \approx 16$ Oe for a kagomé-like lattice coming from a triangular array of l = 400 nm, in agreement with the experimental result. For vortex motion along Y-axis of the array, matching is observed at fractional numbers f = 5/6, 4/6, 3/6, 2/6 but the lowest possible fractional matching field around 16 Oe (f = 1/6 or p = 1 in Table I) is absent. For vortex motion along the X-axis direction the two first fractional minima f = 1/6, 2/6 are missing.

The same qualitative behaviour is also obtained measuring closer to $T_C$, at T = 7.975 K, at different current densities (see Fig. 3): a strong anisotropy between X-axis and Y-axis magnetoresistance curves with several minima at fractional matching conditions below $H^{n=1}$ = 100 Oe. However, as can be seen in Fig. 3, the shape and amplitude of the minima are very sensitive to small changes in the experimental conditions. These subtle changes indicate a complex dynamic behavior in which the dissipation at a particular matching field, current and temperature value depends on the different flow patterns of the vortex lattice.

The observed matching effects in the kagomé-like array are similar to results in other complex lattices such as kagomé and honeycomb [10-12,20-22] in basic aspects: the main matching fields are also given by pinning center density but matching at non integer matching conditions is much more relevant than in simple triangular or square arrays due to the diluted geometry. However, in the present case no matching effects are observed beyond $H^{n=1}$. This is different from honeycomb arrays[20] in which strong matching effects were experimentally observed up to H = 5- 6 times $H^{n=1}$ and from kagomé arrays[10] for fields up to 2 $H^{n=1}$ and in numerical simulations[21] for fields up to 4 $H^{n=1}$. Actually, matching to the pinning center density has also been observed in triangular lattices of antidots, in which some of the pinning centers had been removed at random,[23] but with a strong suppression of higher order matching upon increasing the fraction of missing pinning sites (the second order matching field was found to disappear in the range 20%-40% of antidot dilution, which is comparable to the 33% fraction of missing dots in the kagomé-like array). This lack of commensurability at high fields was attributed to the competition between elastic and pinning energies due to the strong distortions of the vortex lattice needed to match the disordered arrays.[23,24] The same effect should happen in the kagomé-like array. A clear consequence of the interplay between elastic and pinning energies is that, even for the first matching field, the vortex lattice has to extend homogenously for the whole array keeping the appropriate density, so that interstitial or cage vortices placed in the empty triangular areas have to coexist with vortices placed in the triangle-shaped dot areas to assure a homogeneously and uniform vortex lattice. It is interesting to mention that a similar coexistence of local pinning at some regions with interstitial vortices at others has also been found in quasiperiodic arrays of pinning centers,[14] driven by the need to keep constant the global vortex lattice density.

A further indication of the presence of interstitial vortices in the system comes from the anisotropy between Y-axis and X-axis dissipation observed in Figs. 2 and 3. Numerical simulations and experiments in triangular lattices have shown that anisotropic transport is a signature of interstitial vortices that find different channels for vortex flow depending on their propagation direction.[22,25] Thus, the existence of a significant anisotropy in the system, together with its dependence on field and temperature (i.e. vortex density and pinning strength) is, again, a clear signal of the presence of interstitial vortices even at fractional matching conditions.

**3.2 Ratchet effect in kagomé-like arrays.**

Figure 1(c) shows that vortex motion parallel to the Y-axis fulfils the ratchet condition of moving on asymmetric potentials albeit the array is made up of symmetric circular Ni dots. Figure 4 shows the experimental results at temperature close to $T_c$, when alternating driving currents, of 10 kHz frequency, are applied in the X-axis direction, i. e. Lorentz forces and vortex lattice movement in the Y-axis direction, in a similar experimental configuration as in previous works with arrays of Ni triangles.[2] These data show a noticeable rectifier effect created by the global asymmetry of the pinning array interacting with the whole vortex lattice, different from previous results in which the asymmetry was related to the asymmetric shape of the individual pinning centers.[2,4,6,7] The experiments have been done for all the matching field conditions that were analysed in the previous section. There are two main results that should be emphasized: i) the amplitudes of the output dc voltages and the onset of the ratchet effects are very similar for all the matching fields measured, and more striking ii) all the experimental data show vortex ratchet reversal: this reversal is only induced by the strength of the input forces and the number of vortices is playing a minor role in this dilute regime. A reversible rectifier has been obtained which polarity is only tuned by the strength of the ac input signal.

It is worth to mention that, in this kagomé-like array, injecting an ac current parallel to the Y-axis (vortex motion parallel to the X-axis) does not induce any dc output signal, in the same way that was already reported[2,26] for very different asymmetric potentials. Figure 5 shows the ratchet effect when the temperature is decreased, revealing experimental trends similar to Fig. 4 data. The most relevant effect is that, decreasing the temperature, the experimental ratchet threshold increases as has been already reported,[27] i. e., the driving forces, needed to obtain a net vortex flow, increase as the temperature decreases.

Vortex ratchet reversal can be usually related to different vortices moving on two opposite asymmetric potentials[2] which are depicted in the inset of Figure 4 for the kagomé-like array. One of them is the ratchet potential induced by the asymmetric triangular-shaped array of dots and the other is an "empty" cage, a weaker potential, but asymmetric too, with triangular shape and pointing out in the reversed direction. The ratchet effect at low intensity of the driving forces is generated by very mobile vortices which are placed on the "empty" triangles; these vortices are interstitial vortices and they will produce a downward net vortex flow. Increasing the strength of the driving forces, we can reach the threshold force for vortices placed on the triangular-shaped array of dots; these vortices move upward and the polarity of the dc voltage changes. However, the independence of ratchet reversal on the order of the matching field observed in this kagomé-like array utterly deviates from the previously reported behaviour of ratchet effects and vortex reversal in arrays of Ni triangles.[2] In that case, the number of vortices per pinning center (n) was crucial to yield vortex ratchet reversal with a threshold at n=3, the filling factor of the Ni triangles while in the present case ratchet reversal is found even for n <1. Thus, the observed ratchet reversal here would imply that for every one of the matching fields, mobile and weakly pinned interstitial and less mobile and more pinned vortices coexist. This different behavior can be directly linked to the centers that create the asymmetric potential in each case; a full Ni triangle in Ref. 2 and a triangular arrangement of Ni dots here (see inset Fig 4). As was discussed in the previous section 3.1, the observed matching fields correspond to vortices in the triangular arrangement of Ni dots. However, this implies an inhomogeneous density in the sample. In the kagomé-like lattice the competition between the vortex-vortex and vortex-pinning interactions will favour homogeneous vortex density. This competition would not be so relevant in the case of full Ni triangles, since all the pinned vortices are located in a region of depressed superconductivity.

Thus, two key factors appear to obtain a rectifier in which the polarity of the ratchet effect can just be tuned by the strength of driving forces independent of the matching field: i) the asymmetry in the global lattice (to induce the asymmetry in the pinning potential) and ii) the existence of a regular distribution of areas with and without pinning centers, so that the interplay between elastic and pinning energies favours the existence of interstitial vortices at all matching fields, since, as was discussed in the previous section, the vortex lattice has to cover homogeneously the defined pattern keeping the appropriate density.

## 4. CONCLUSIONS

Vortex lattice dynamics on kagomé-like periodic pinning centers provides a rich scenario for the study of matching and ratchet effects. In this array an asymmetric pinning potential is produced by the asymmetric arrangement of circular Ni dots: triangular arrangements of six Ni dots constitute the basic pinning areas, intercalated with empty triangular regions pointing in the opposite direction. Synchronized vortex lattice motion is observed at the first integer matching field and at several fractional matching fields, which correspond to smaller magnetic fields than the first (main) matching field value. The former happens when the density of vortices matches the density of pinning centers. The fractional ones occur when the lattice vortex density matches lower density than the actual pinning centers density. These vortex densities correspond to the densities obtained taking out one by one pinning centers.

Ratchet effects are observed when an ac current is injected perpendicular to the reflection symmetry axis: the kagomé-like array behaves as a reversible rectifier in which the polarity can be tuned by the strength of the driving forces independently of the matching field. This remarkable behavior is the result of the existence of interstitial vortices at all matching fields due to the competition between intervortex interaction that favours a homogeneous vortex lattice and pinning by the distribution of Ni dots in the kagomé-like array.

**Acknowledgments**

We want to thank useful conversations with Ivan K. Schuller. We want to acknowledge the support by Spanish Ministerio Ciencia e Innovación grants FIS2008-06249 (Grupo Consolidado), Consolider CSD2007-00010 and HP2008-0032, CAM grant S2009/MAT-1726, Santander-UCM grant GR58/08, Principado de Asturias FICYT grant IB08-106 and IMDEA-Nanoscience.

**References**

[1] M. Vélez, J.I. Martín, J. E. Villegas, A. Hoffmann, E. M. González, J. L. Vicent and I. K. Schuller, J. Magn. Magn. Mat. **320**, 2547 (2008).

[2] J. E. Villegas, S. Savel'ev, F. Nori, E. M. González, J. V. Anguita, R. Garcia, and J. L. Vicent, Science **302**, 1188 (2003).

[3] P. Hänggi and F. Marchesoni, Rev. Mod. Phys. **81**, 387 (2009).

[4] C. C. de Souza Silva, J. Van de Vondel, M. Morelle, and V. V. Moshchalkov, Nature **440**, 651 (2006).

[5] W. Gillijns, A.V. Silhanek, V. V. Moshchalkov, C. J. O. Reichhardt and C. Reichhardt, Phys. Rev. Lett. **99**, 247002 (2007).

[6] T. C. Wu, L. Horng, J. C. Wu, R. Cao, J. Kolacek, and T. J. Yang, J. Appl. Phys. **102**, 033918 (2007).

[7] Y. Togawa, K. Harada, T. Akashi, H. Kasai, T. Matsuda, F. Nori, A. Maeda, and A. Tonomura, Phys. Rev. Lett. **95**, 087002 (2005).

[8] O. M. Stoll, M. I. Montero, J. Guimpel, J. J. Akerman, and I. K. Schuller, Phys. Rev. B **65,** 104518 (2002).

[9] J. I. Martin, M. Velez, A. Hoffmann, I. K. Schuller, and J. L. Vicent Phys. Rev. Lett. **83**, 1022 (1999).

[10] D. J. Morgan, and J. B. Ketterson, Phys.Rev. Lett. **80**, 3614 (1998).

[11] M. F. Laguna, C. A. Balseiro, D. Dominguez, and F. Nori, Phys. Rev. B **64,** 104505 (2001).

[12] T. C. Wu, J. C. Wang, L. Horng, J. C. Wu, and T. J. Yang, J. Appl. Phys. **97**, 10B102 (2005).

[13] C. Reichhardt, and C. J. Olson Reichhardt, Phys. Rev. B **81,** 024510 (2010).

[14] J. E. Villegas, M. I. Montero, C. P. Li, and I. K. Schuller, Phys. Rev. Lett. **97**, 027002 (2006).

[15] Y. J. Rosen, A. Sharoni, and I. K. Schuller, Phys. Rev. B **82,** 014509 (2010).

[16] J. I. Martin, M. Velez, J. Nogues, and I. K. Schuller, Phys. Rev. Lett. **79**, 1929 (1997).

[17] M. Baert, V. V. Metlushko, R. Jonckheere, V. V. Moshchalkov, and Y. Bruynseraede, Phys. Rev. Lett. **74**, 3269 (1995).

[18] E. Rosseel, M. Van Bael, M. Baert, R. Jonckheere, V. V. Moshchalkov, and Y. Bruynseraede, Phys. Rev. B **53,** R2983 (1996).

[19] J. I. Martin, Y. Jaccard, A. Hoffmann, J. Nogues, J. M. George, J. L. Vicent, and I. K. Schuller, J. Appl. Phys. **84**, 411 (1998).

[20] R. Cao, L. Horng, T. C. Wu and T. J. Yang, J. Phys.: Condens. Matter. **21**, 075705 (2009).

[21] C. Reichhardt and C. J. Olson Reichhardt, Phys. Rev. B **76**, 064523 (2007).

[22] C. Reichhardt and C. J. Olson Reichhardt, Phys. Rev. B **79**, 134501 (2009).

[23] M. Kemmler, D. Bothner, K. Ilin, M. Siegel, R. Kleiner and D. Koelle, Phys. Rev. B **79**, 184509 (2009).

[24] C. Reichhardt and C. J. Olson Reichhardt, Phys. Rev. B **76**, 094512 (2007).

[25] R. Cao, T. C. Wu, P. C. Kang, J. C. Wu, T. J. Yang, and L. Horng, Solid State Commun. **143**, 171 (2007).

[26] D. Perez de Lara, F. J. Castaño, B. G. Ng, H. S. Kröner, R. K. Dumas, E. M. Gonzalez, Kai Liu, C. A. Ross, Ivan K. Schuller, and J. L. Vicent, Phys. Rev. **B80**, 224510 (2009).

[27] J. E. Villegas, E. M. Gonzalez, M. P. Gonzalez, J. V. Anguita, and J. L. Vicent, Phys. Rev. **B71**, 024519 (2005).

**Figure Captions**

Figure 1. Sketch of different arrays of pinning centers. (a) Triangular array. (b) Kagomé array. (c) Kagomé-like array. The lines and shadow areas show how the arrays are constructed. That is, the kagomé array can be derived from the triangular one by eliminating 1/4 of the pining sites. In the case of our kagomé-like array, a fraction of 3/9 are eliminated from the triangular one, thus only a fraction of 6/9 remains forming a triangular building block.

Figure 2. Resistance *vs* applied magnetic field of a hybrid Nb film on kagomé-like array of Ni dots at temperature T=7.95 K ($T_c$ = 8.03 K).
(a) Current density $J_{DC}$ = 2.0 × 10$^8$ A m$^{-2}$ injected parallel to the X- axis of the array, vortex motion parallel to the Y-axis. Inset: Magnetic field position of the resistance minima vs. index number p for Y-axis direction of vortex motion. The correlation coefficient of the linear fit is 0.997. Solid lines are linear fit to the data
(b) Current density $J_{DC}$ = 1.9 × 10$^8$ A m$^{-2}$ injected parallel to the Y- axis of the array, vortex motion parallel to the X-axis. Inset: Magnetic field position of the resistance minima vs. index number p for X-axis direction of vortex motion. The correlation coefficient of the linear fit is 0.995. Solid lines are linear fit to the data

Figure 3. Resistance *vs* applied magnetic field of a hybrid Nb film on kagomé-like array of Ni dots at temperature T=7.975 K ($T_c$ = 8.03 K). (a) Current density $J_{DC}$ applied parallel to the X- axis of the array (see Fig. 1 (c)), vortex motion parallel to the Y-axis. $J_{DC}$ = 2 × 10$^8$ A m$^{-2}$ (upper curve), and $J_{DC}$ = 1.6 × 10$^8$ A m$^{-2}$ (lower curve). (b) Current density $J_{DC}$ applied parallel to the Y- axis of the array (see Fig. 1 (c)), vortex motion parallel to the X-axis. $J_{DC}$ = 2 × 10$^8$ A m$^{-2}$ (upper curve), and $J_{DC}$ = 1 × 10$^8$ A m$^{-2}$ (lower curve).

Figure 4. Ratchet effect measurements in a hybrid Nb film on kagomé-like array of Ni dots at T=7.95 K ($T_c$ = 8.03 K). Frequency of the input ac current is 10 kHz. The graphs show the dc output voltages *vs* the ac input current densities at different matching fields.

Inset: Kagomé-like array showing the two asymmetric triangular pinning wells (see text).

Figure 5. Ratchet effect measurements in a hybrid Nb film on kagomé-like array of Ni dots at T=7.87 K ($T_c$ = 8.03 K). Frequency of the ac input current is 10 kHz. The graphs show the dc output voltages *vs* the ac input current densities at different matching fields.

**Table Caption**

Table I. Theoretical positions of minima and experimental values for the first integer matching field (labelled p=6 or n=1) and fractional minima. The fractional minima are labelled with p, being p defined as the number of pining sites of 6 that remain to calculate the density of vortex lattice in this kagomé-like arrangement. Then, the fraction f of pining sites is defined as f=p/6. In the table, for each value of p, we show in rows below: first, the calculated position of minima according to the description given in the text; second, the direction of vortex lattice motion X and Y as shown in Fig. 1(c) and, finally, the observed experimental position of the minima for positive and negative fields.

| f = p/6 (l = 400 nm) | p=6 (n=1) | | p=5 | | p=4 | | p=3 | | p=2 | p=1 |
|---|---|---|---|---|---|---|---|---|---|---|
| Theoretical Position (Oe) | 99.4 | | 82.8 | | 66.2 | | 49.7 | | 33.1 | 16.5 |
| Vortex motion direction | Y | X | Y | X | Y | X | Y | X | Y | |
| Experimental position (Oe) Positive Fields | 99 | 101 | 81 | 86 | 66 | 63 | 47 | 40 | 25 | |
| Experimental position (Oe) Negative Fields | -99 | -100 | -80 | -85 | -68 | -61 | -51 | -43 | -27 | |

Fig. 1

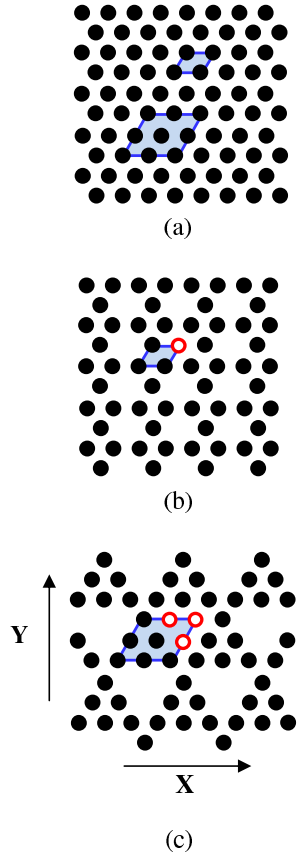

(a)

(b)

(c)

Figure 1      BV10965      14OCT2010

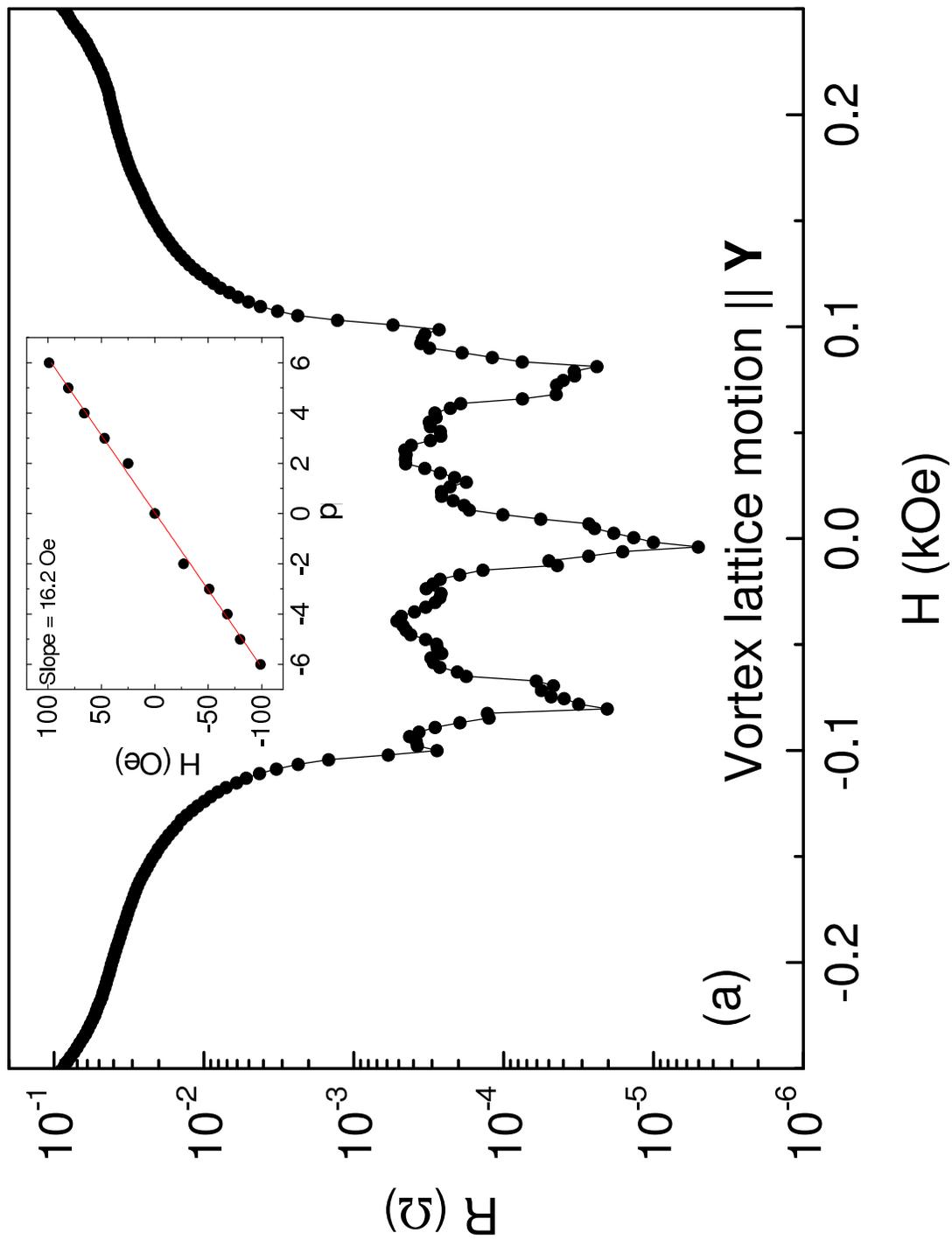

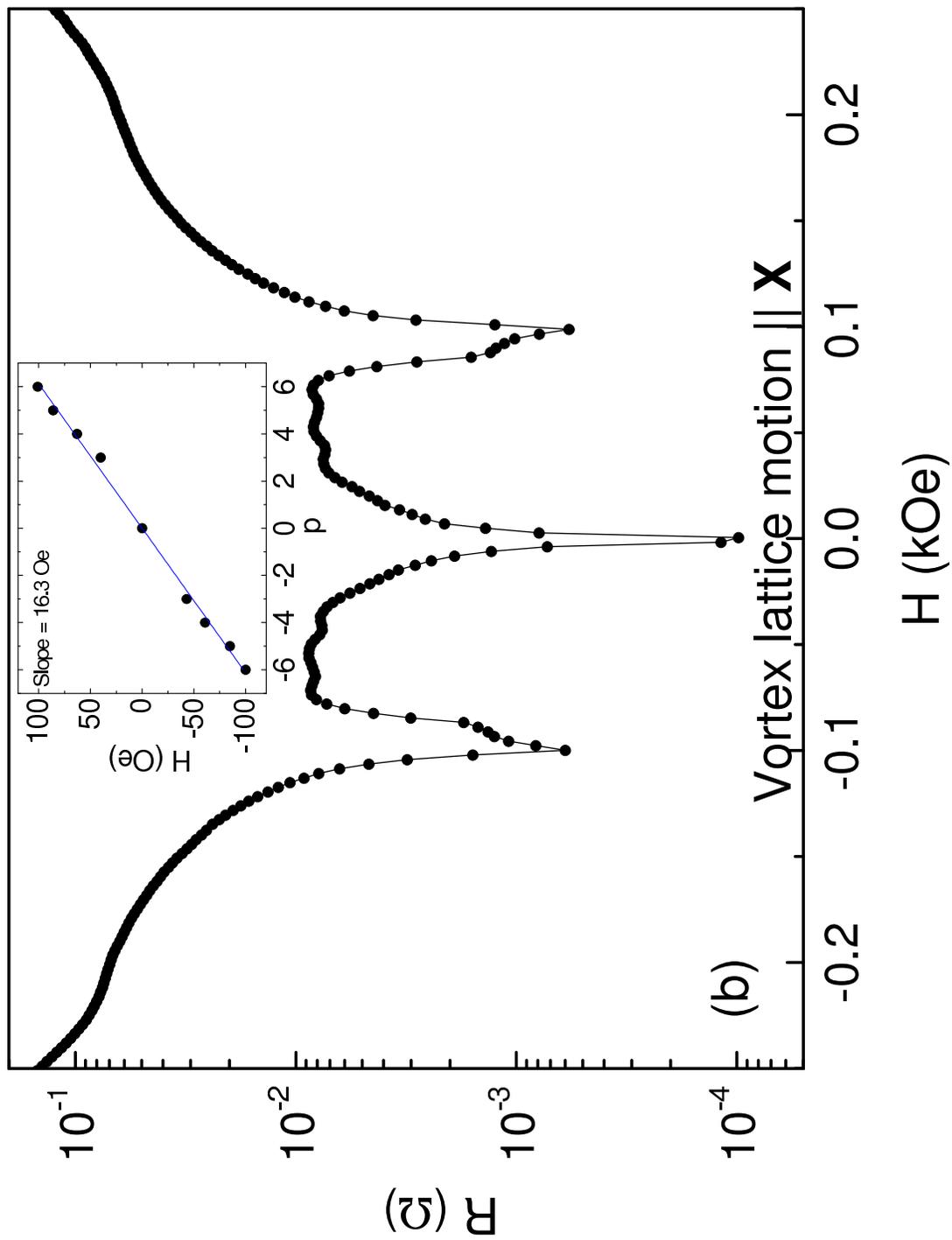

Figure 2b    BV10965    14OCT2010

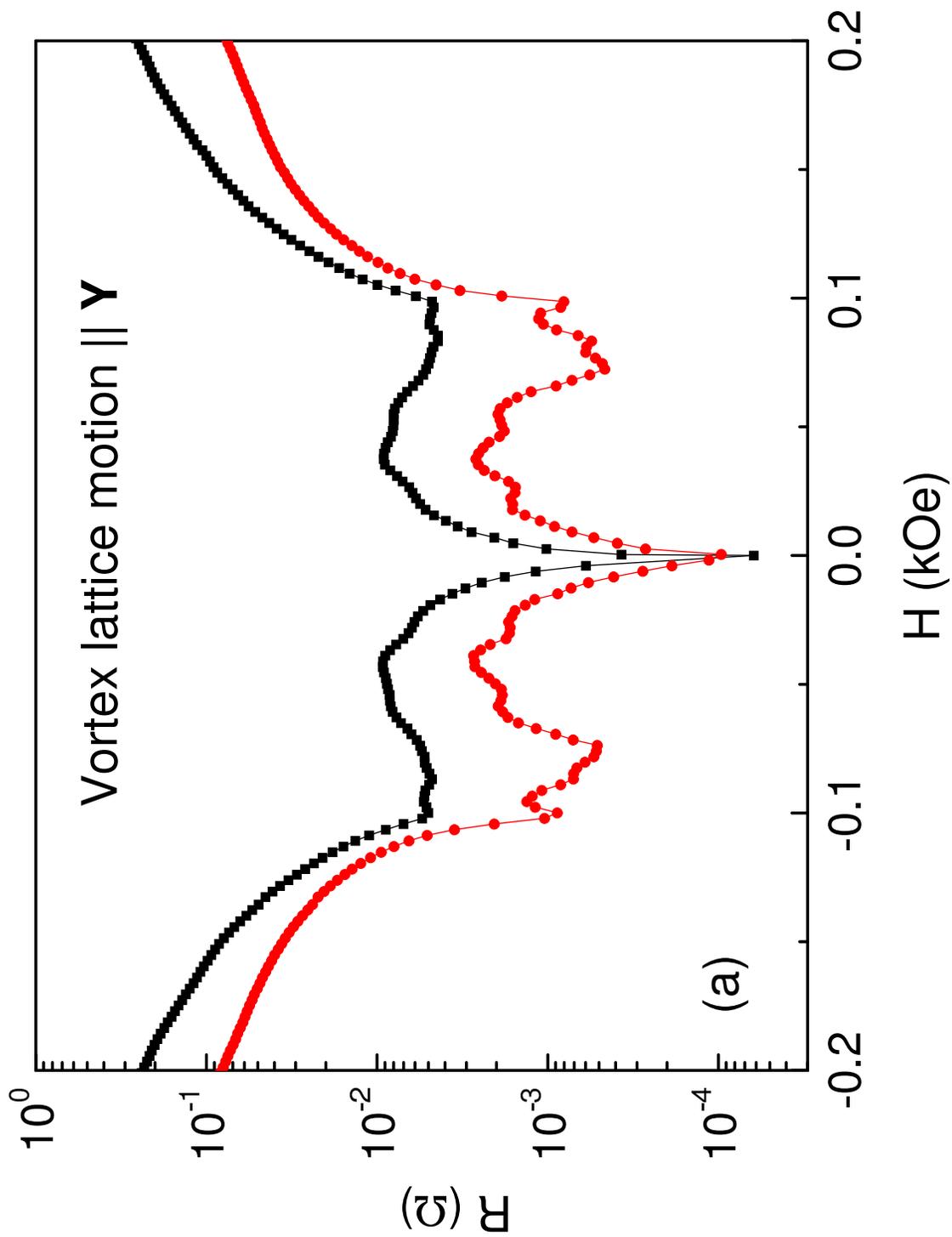

Figure 3a     BV10965     14OCT2010

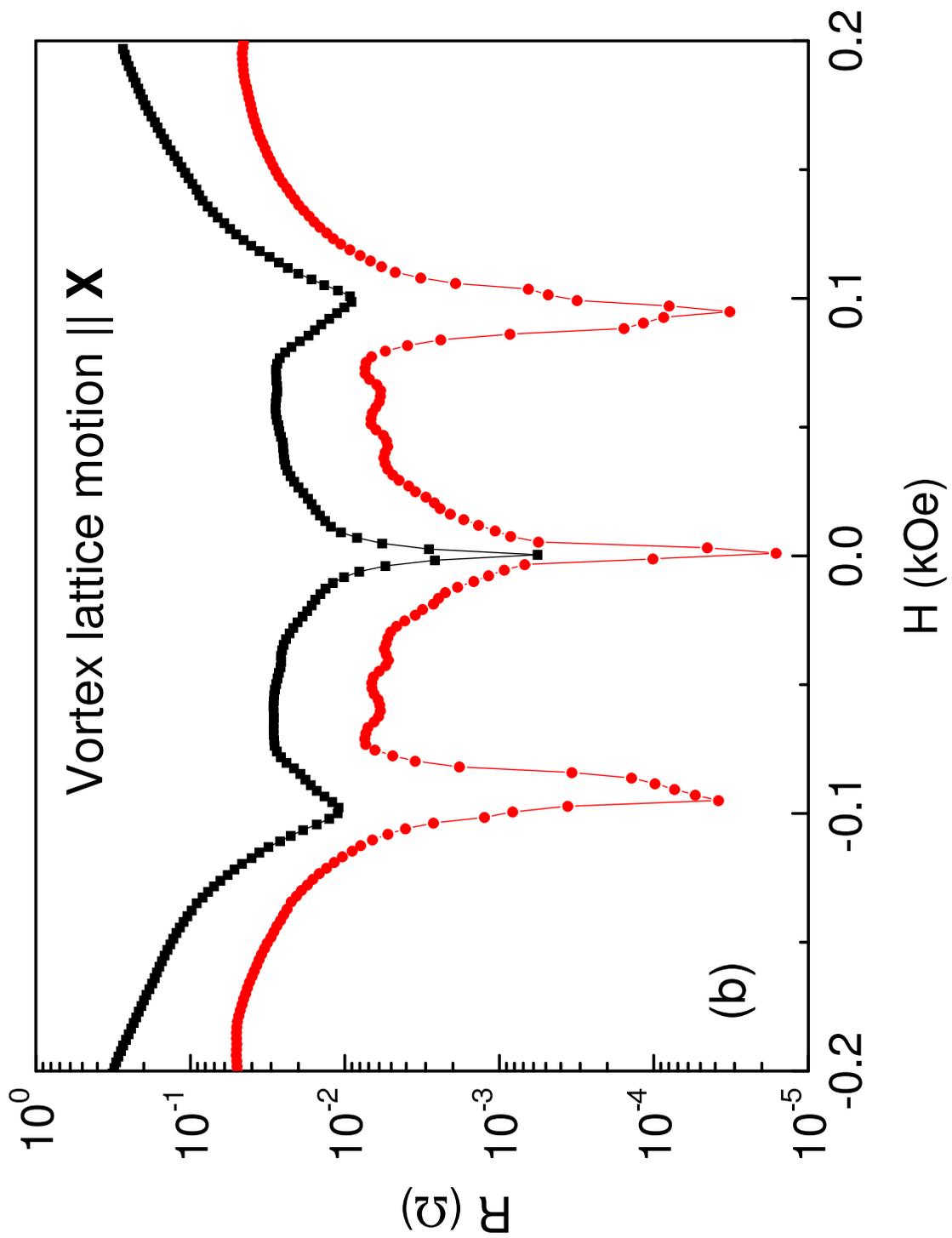

Figure 3b    BV10965    14OCT2010

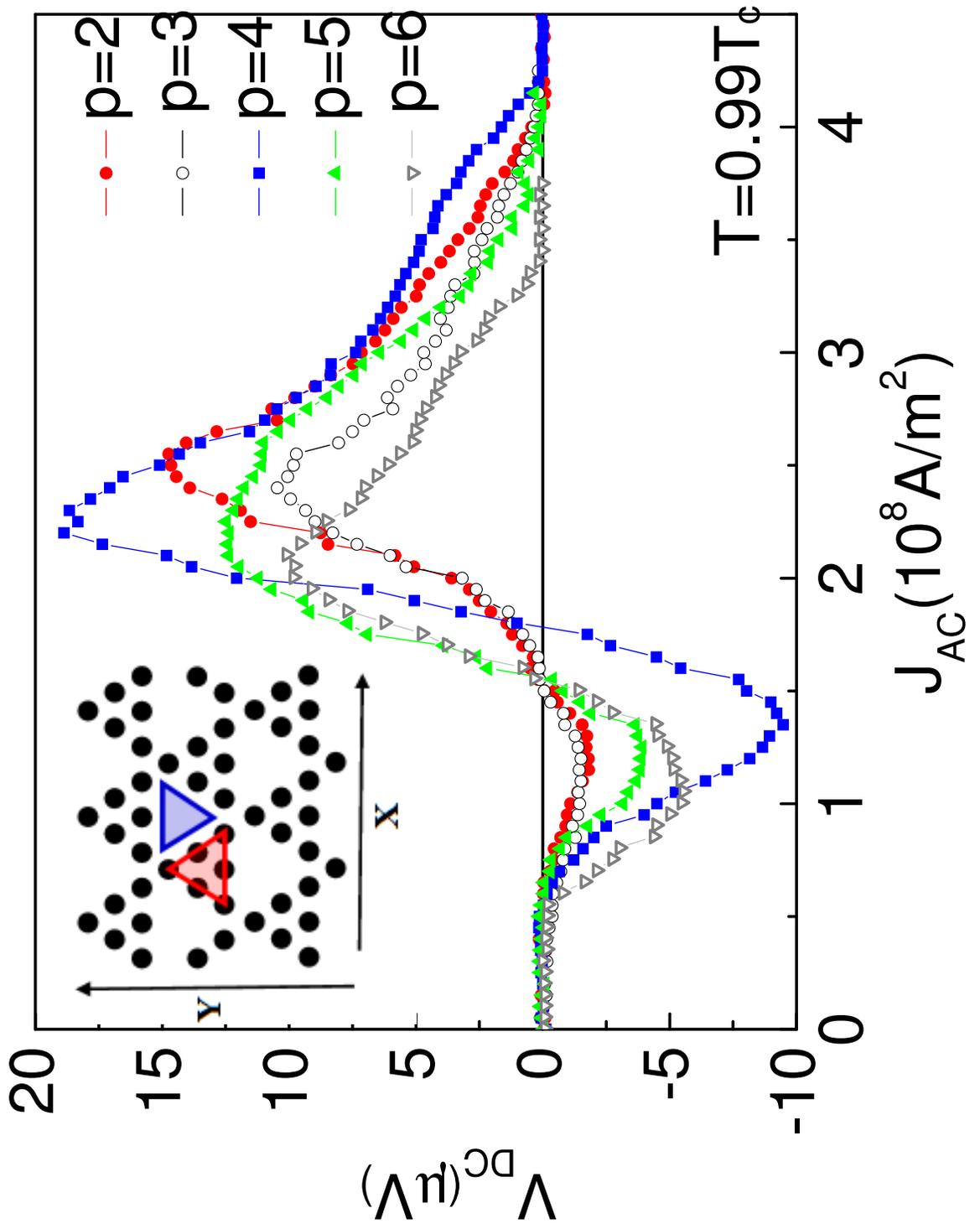

Figure 4 BV10965 14OCT2010

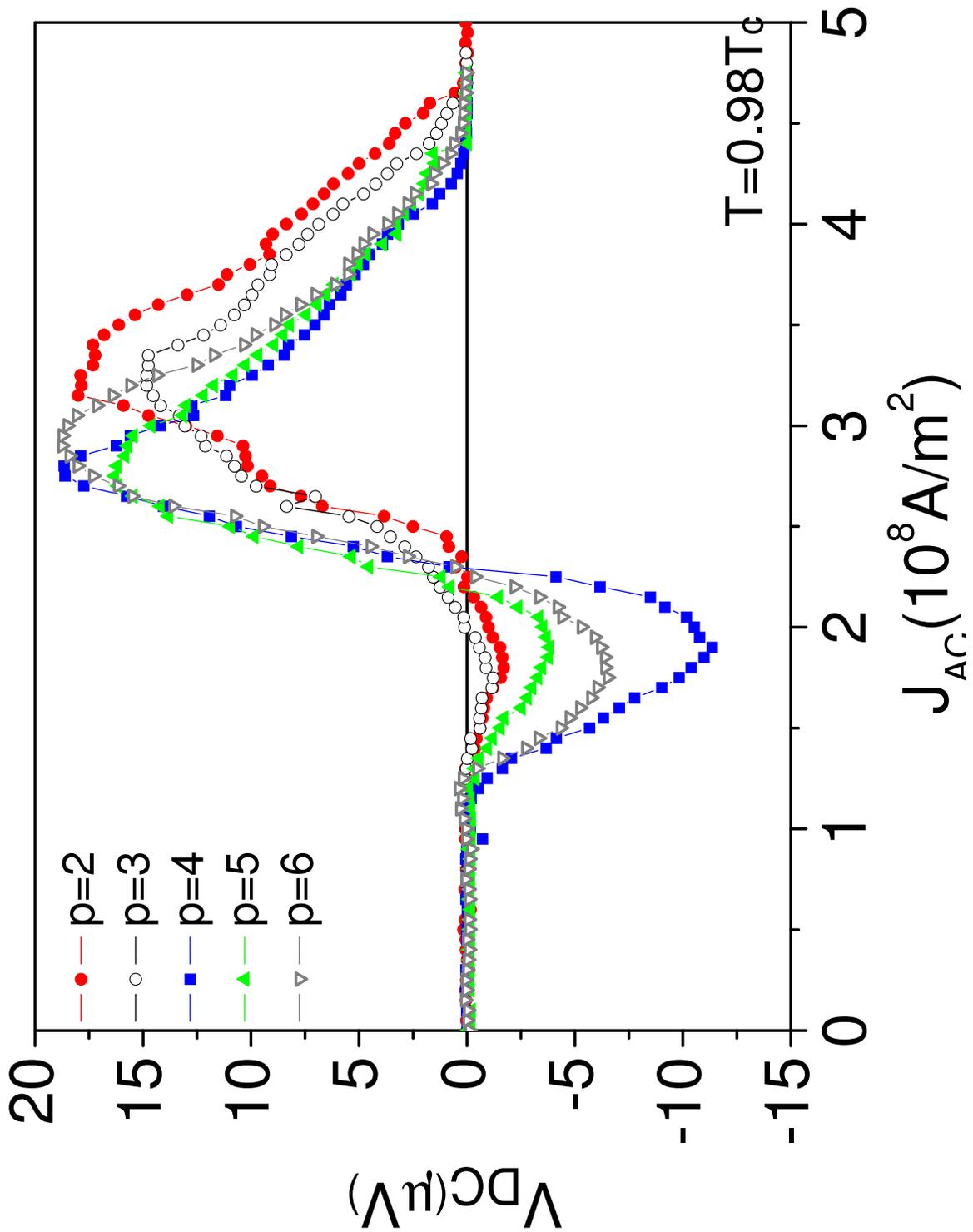

Figure 5        BV10965     14OCT2010